\documentclass[%
reprint,
 amsmath,amssymb,
prl,
]{revtex4-2}

\usepackage{graphicx}
\usepackage{dcolumn}
\usepackage{bm}
\usepackage{hyperref}
 \usepackage{xcolor}

\begin{document}


\title{The Available Energy of Trapped Electrons and Its Relation to Turbulent Transport}

\author{R.J.J. Mackenbach$^{1,2}$} 
\email{r.j.j.mackenbach@tue.nl}
\author{J.H.E. Proll$^1$}
\author{P. Helander$^2$}
\affiliation{$^1$Eindhoven University of Technology, 5612 AZ, Eindhoven, Netherlands \\
$^2$Max Planck Institute for Plasma Physics, 17491 Greifswald, Germany}

\date{\today}
\begin{abstract}
Any collisionless plasma possesses some ``available energy'' (AE), defined as that part of the thermal energy that can be converted into instabilities and turbulence. Here, we present a calculation of the AE carried by magnetically trapped electrons in a flux tube of collisionless plasma. The AE is compared with nonlinear simulations of the energy flux resulting from collisionless turbulence driven by trapped-electron modes in various magnetic geometries. The numerical calculation of AE is rapid and shows a strong correlation with the simulated energy fluxes, which can be expressed as a power law and understood in terms of a simple model. 
\end{abstract}

\maketitle


\textit{Introduction.}
One of the greatest difficulties facing magnetic-confinement fusion is the degradation of energy confinement due to plasma turbulence. Tokamaks and stellarators suffer from substantial turbulent energy losses, which effectively set a lower limit on the size and/or magnetic field strength of these devices. Turbulence arises because the density and temperature vary over the plasma volume, but the level of turbulence also depends on the geometry of the magnetic field in ways that are not completely understood. Over the years, an enormous amount of work has been devoted to the identification of linear plasma instabilities and the construction of gyrokinetic codes for simulating plasma turbulence, but the outcome of such simulations is often difficult to predict or interpret. For instance, when different plasma configurations are compared, the turbulent energy flux often differs much more than the linear instability growth rates. 
\par
In the present Letter, we propose a nonlinear measure for predicting the strength of turbulence, which is computationally efficient and does not rely on direct simulation. The measure is based on the concept of available energy (AE), which quantifies how much energy can, in principle, be converted into turbulent motion. This notion was first introduced in meteorology by Lorenz \cite{Lorenz1955AvailableCirculation}, and in plasma physics by Gardner \cite{Gardner1963BoundPlasma}. Lorenz found that less than one percent of the potential energy of the atmosphere is generally available for conversion into kinetic energy, and Gardner noted that the AE of a collisionless Vlasov plasma is severely restricted by constraints imposed by Liouville's theorem. The concept of AE has since found use in varying fields where the dynamics obey Liouville's theorem, ranging from galactic clusters to Bose-Einstein condensates \cite{Berk1970PhaseObservations,Bartholomew1971OnGalaxies,Stahl1998ASystems,Chavanis2012DynamicalMethod,Lemou2012OrbitalModels,Chavanis2012DynamicalMethod,Baldovin2016NonequilibriumGases}. This utility also extends to systems with diffusion, as implied by a recent proof by Kolmes and Fisch \cite{Kolmes2020RecoveringOperations}. \par  
In a magnetically confined plasma, the AE is further limited by adiabatic invariants  \cite{Helander2017AvailablePlasmas,Helander2020AvailablePlasmas}, which, importantly, cause it to depend on the geometry of the magnetic field. Focusing on turbulence resulting from the collisionless trapped-electron mode (TEM) \cite{Kadomtsev1967PlasmaGeometry,Dannert2005GyrokineticTurbulence}, driven either by a density or an electron temperature gradient, we suggest that the observed sensitivity of TEM turbulence to magnetic-field geometry in tokamaks and stellarators can be understood in terms of the AE. 
\par
For a collisionless plasma the AE is defined as the difference in thermal energy between the ``initial'' distribution function $f_{\mathrm{ini}}$, of which one wishes to calculate the AE, and that of the so-called ground-state $f_g$. The latter is defined as the distribution function that minimizes the thermal energy, subject to constraints imposed by Liouville's theorem and adiabatic invariants. In recent work \cite{Helander2020AvailablePlasmas}, it was found that the ground state of trapped electrons is a function of the particle energy $\epsilon$, the magnetic moment $\mu$, and the second adiabatic invariant $$\mathcal{J} = \int m v_\| \, d \ell $$ 
alone. This and all similar integrals below are taken along the magnetic field between two consecutive bounce points of the particle trajectory. The mass is denoted by $m$, the velocity along the magnetic field by $v_\|$, and the arc length in this direction by $\ell$. Furthermore, the ground state was found to be a decreasing function of $\epsilon$ for all values of $\mu$ and $\mathcal{J}$.
\par 
Using these results, an expression for the AE of trapped electrons was derived by employing a coordinate system consisting of the \textit{toroidal} flux $\psi$ (some authors reserve this symbol for the poloidal flux) and the Clebsch angle $\alpha$, which locally define the magnetic field as $\mathbf{B} = \nabla \psi \times \nabla \alpha$ \cite{Dhaeseleer2012FluxTheory}. In these coordinates, the ground state $f_g(\epsilon,\mu,\cal J)$ obeys the integrodifferential equation
\begin{equation}
    \begin{aligned}
        \frac{ \partial f_g(w,\mu,\mathcal{J})}{\partial w}  = - \frac{\iint \delta [w - \epsilon(\psi,\alpha,\mu,\mathcal{J})] \, d\psi d\alpha }{\iint \delta[f_{\mathrm{ini}} - f_g(w,\mu,\mathcal{J})] \, d\psi d\alpha },
    \end{aligned}
\label{eq:ground-state-eq}
\end{equation}
where $\delta[x]$ denotes the Dirac-delta distribution, and $w$ is a positive scalar.
From this equation, the AE can be computed for the case of an omnigenous flux tube, in which the parallel invariant $\mathcal{J}$ is independent of the Clebsch angle $\alpha$ for all electron orbits \cite{Hall1975Three-dimensionalPlasma}. 
This condition is however not satisfied in most stellarators, and we therefore extend the derivation to nonomnigenous systems in order to be able to compare the result with nonlinear turbulence simulations. In doing so, we find that the AE correlates closely with the nonlinear energy flux in turbulence simulations by the gyrokinetic code \textsc{gene} \cite{Jenko2000ElectronTurbulence} over several orders of magnitude, across a tokamak and various stellarator devices. The transport is found to follow a simple power law in AE, which can be motivated by a simple argument.

\textit{Theory.}
In the calculation of the ground state and AE, we restrict our attention to a subregion $\Omega$ of the plasma in the shape of a slender flux tube \cite{Beer1995FieldTurbulence}, which allows us to approximate the distribution function by its first-order Taylor expansion in the directions perpendicular to the magnetic field. The calculation is particularly simple if the cross section of the flux tube is elliptical in the $(\psi,\alpha)$ plane, so that $\Omega$ corresponds to 
\begin{equation*}
    \Big( \frac{\psi - \psi_0}{\Delta \psi} \Big)^2 + \Big( \frac{\alpha - \alpha_0}{\Delta \alpha} \Big)^2 \leq 1.
    \label{eq:domain-shape}
\end{equation*}
Here $\psi_0$ and $\alpha_0$ are the flux-surface label and field-line label of the magnetic field line defining the center of the flux tube, and $\Delta \psi$ and $\Delta \alpha$ define its width in the $\psi$ and $\alpha$ directions, respectively. There is no reason to suspect that the AE is very different in flux tubes with other cross sections, but choosing different cross sections greatly obfuscates the calculation of the AE. Gyrokinetic simulations are usually carried out in flux tubes with rectangular cross section in the $(\psi,\alpha)$ plane, but will nevertheless be compared with our analytical expressions below.
\par 
In order to find the ground state as in Eq. \eqref{eq:ground-state-eq}, one needs to evaluate integrals of the form
\begin{equation*}
    I[h] \equiv \int_\Omega \delta[h] \, \mathrm{d}\psi \mathrm{d}\alpha,
\end{equation*}
where $h(\psi,\alpha)$ is an arbitrary smooth function that vanishes when $(\psi,\alpha)=(\psi_0,\alpha_0)$. To leading order in $\Delta \psi$ and $\Delta \alpha$, this integral reduces to \cite{Hormander2015TheAnalysis}
\begin{equation*}
    I[h] = \frac{2 \Delta \psi \Delta \alpha}{\sqrt{( \Delta \psi )^2 ( \partial h/\partial \psi )^2 + (\Delta \alpha )^2 ( \partial h/\partial \alpha )^2}}.
    \label{eq:integral-eq-for-ground-state}
\end{equation*}
We now have sufficient information to solve Eq. \eqref{eq:ground-state-eq} to the requisite accuracy. We take the initial distribution function to be a Maxwellian,
\begin{equation}
    f_{\mathrm{ini}} = f_M = n(\psi) \left( \frac{m}{2 \pi T(\psi)} \right)^{3/2} \exp{\left( - \frac{\epsilon}{T(\psi)} \right) },
\end{equation}
where $T(\psi)$ denotes the electron temperature and $n(\psi)$ the electron number density. One can further relate the various derivatives of the particle energy $\epsilon$ to bounce-averaged frequencies \cite{Rosenbluth1971FiniteInstability,Helander2020AvailablePlasmas,Helander2014TheoryFields}, giving
\begin{subequations}
    \begin{alignat}{2}
        \left(\frac{\partial \epsilon}{\partial \psi}\right)_{\mu,\mathcal{J},\alpha} &= - && e \omega_\alpha, \\
    \left(\frac{\partial \epsilon}{\partial \alpha}\right)_{\mu,\mathcal{J},\psi} &=  && e \omega_\psi,
    \end{alignat}
\end{subequations}
with $e$ being the elementary charge. We have introduced two bounce-averaged drift frequencies, which are equal to
\begin{subequations}
    \begin{alignat}{2}
        \omega_\alpha &= \frac{1}{\tau_{b}} \int \left(\mathbf{v}_{d} \cdot \nabla \alpha\right) \frac{d \ell}{v_{\|}} 
        =  & \frac{1}{e \tau_b} \left( \frac{\partial {\cal J}}{\partial \psi} \right)_{\epsilon,\mu,\alpha}, \\
        \omega_\psi &=  \frac{1}{\tau_{b}} \int \left(\mathbf{v}_{d} \cdot \nabla \psi \right) \frac{d \ell}{v_{\|}}
        = - &\frac{1}{e \tau_b} \left( \frac{\partial {\cal J}}{\partial \alpha} \right)_{\epsilon,\mu,\psi},\\
        \tau_b &= \int \frac{d \ell}{v_{\|}}
        = \left( \frac{\partial {\cal J}}{\partial \epsilon} \right)_{\mu,\psi,\alpha}. &
    \end{alignat}
\end{subequations}
Here $\mathbf{v}_d$ is the guiding-center drift velocity, and $\tau_b$ is the time it takes for a trapped particle to travel between two consecutive bounce points. 
Using these results, we find that the ground state is given by
\begin{subequations}
    \begin{alignat}{1}
        \label{eq:ground-state-final}
        \frac{ \partial f_g}{\partial w} &= - \frac{f_{M0}}{T_0} F, \\
        \label{eq:non-linear-part-of-ground-state}
        F & \equiv \frac{\sqrt{(\omega_*^T - \omega_\alpha)^2 (\Delta \psi)^2 + \omega_\psi^2 (\Delta \alpha)^2}}{\sqrt{\omega_\alpha^2 (\Delta \psi)^2 + \omega_\psi^2 (\Delta \alpha)^2} }.
    \end{alignat}
\end{subequations}
Here, quantities with a subscript zero are evaluated at the center of the flux tube, i.e., $\psi = \psi_0$ and $\alpha = \alpha_0$. Furthermore we have introduced the diamagnetic frequency $\omega_*^T$, which depends on the gradients of the temperature and number density, in the following manner:
\begin{equation}
    \omega_*^T = -\frac{T_0}{e} \left( \frac{\partial \ln n}{\partial \psi} + \frac{\partial \ln T}{\partial \psi} \left[ \frac{\epsilon_0}{T_0} - \frac{3}{2} \right] \right).
\end{equation}
\par 
To evaluate the AE, one needs to find the difference in energy between the Maxwellian and the ground state to leading order in $\Delta \psi$ and $\Delta \alpha$ by computing the integral
\begin{equation*}
\begin{aligned}
    A = \int \epsilon \left( f_M - f_g  \right) d \mathbf{x},
\end{aligned}
\end{equation*}
taken over all phase-space coordinates $\mathbf{x}$. The expansion of this integral has previously been evaluated in Ref.~\cite{Helander2020AvailablePlasmas}, and following this methodology we find
\begin{equation}
    \begin{aligned}
          A = & \pi^2 \left( \frac{ e \Delta \psi \Delta \alpha }{m}\right)^2 \iint  d\mu d\mathcal{J} ~ \frac{f_{M0}}{T_0} \times \\
          & \Bigg[  \omega_\alpha^2 \left( \frac{\omega_*^T}{\omega_\alpha} - 1 +  F \right) \frac{\Delta \psi}{\Delta \alpha} + \omega_\psi^2 \left( -1 + F \right) \frac{\Delta \alpha}{\Delta \psi} \Bigg].
    \end{aligned}
    \label{eq:AE-dimensionfull-form}
\end{equation} 
Here, the integrand is positive definite, and so is therefore the AE, as it must be by definition. Taking the limit of omnigeneity ($\omega_\psi = 0$) in this expression, one retrieves the result previously found by Helander \cite{Helander2020AvailablePlasmas}, according to which the plasma is stable to electron-driven modes if $\omega_*^T / \omega_\alpha \leq 1$ for all particle orbits. This is a well-established result from linear theory \cite{Proll2012ResilienceInstabilities}, which also holds nonlinearly \cite{Helander2017AvailablePlasmas}.
\par
Our next step is to relate the AE to typical turbulence quantities, but there is a basic difficulty having to do with the question of how these scale with the system size. Some types of turbulence depend on the size and shape of the domain in which it takes place, but we are mainly interested in turbulence for which this is not the case if the domain is large enough. We shall refer to such turbulence as ``local.'' 
If we expect AE to encapsulate information about local turbulence, it should only refer to the energy available over some characteristic length scale comparable to the correlation length of the turbulent fluctuations. We denote this length by $\Delta \psi_A$ and $\Delta \alpha_A$, and assume that it is of the order of the poloidal gyroradius, as is usually observed in gyrokinetic simulations, and since it is the poloidal flux which is mostly responsible for confinement in tokamaks and stellarators.
\par

We are now in a position to define a dimensionless AE which will be used in the rest of the analysis. As in neoclassical transport theory \citep{Helander2005CollisionalPlasmas}, it is useful to perform a change of variables, $(\mu,\mathcal{J}) \mapsto (\lambda, z)$, with
\[
    \lambda = \frac{\mu \overline{B}}{\epsilon_0}, \quad z = \frac{\epsilon_0}{T_0}, 
\]
where $\overline{B}$ is the average magnetic field strength. Employing these new integration coordinates, we find that the available energy becomes
\[ 
A = \left( \frac{n_0 T_0}{4\sqrt{\pi}} \frac{\Delta \psi_A \Delta \alpha_A L}{\overline{B}} \right) \widehat{A},
\]
where $L$ is the total length of the magnetic field line, and the factor in brackets is thus proportional to the total thermal energy in the domain. The dimensionless factor $\widehat{A}$ is defined as
\begin{equation}
    \begin{aligned}
        \widehat{A} \equiv & \iint dzd\lambda \sum_{\text{wells}(\lambda)} \exp(-z) z^{5/2} \times \\ 
        & \left[ \hat{\omega}_{\alpha}^2 \left( \frac{ \hat{\omega}_{*}^T }{\hat{\omega}_{\alpha}} - 1 +  \hat{F} \right) + \hat{\omega}_{\psi}^2 \left( -1 + \hat{F} \right)  \right] \hat{G}^{1/2},
    \end{aligned}
    \label{eq:normalized-AE-final}
\end{equation}
where the summation over  wells$(\lambda)$ is taken over all magnetic trapping wells along the flux tube associated with a specific value of $\lambda$. Several dimensionless variables have been introduced here. Firstly, the dimensionless precession frequencies are defined as
\begin{equation*}
    \begin{aligned}
        \hat{\omega}_\alpha \equiv \frac{e \Delta \psi_A}{\epsilon_0} \omega_\alpha, \quad
        \hat{\omega}_\psi   \equiv \frac{e \Delta \alpha_A}{\epsilon_0} \omega_\psi, \quad
        \hat{\omega}_*^T    \equiv \frac{e \Delta \psi_A}{\epsilon_0} \omega_*^T.
    \end{aligned}
\end{equation*}
Secondly, the dimensionless Jacobian $\hat{G}^{1/2}$ is defined as
\begin{equation*}
    \hat{G}^{1/2} \equiv \frac{\tau_b}{L} \sqrt{\frac{2 \epsilon_0}{m}}.
\end{equation*}
The nonlinear function $\hat{F}$ is defined as
\begin{equation*}
    \hat{F} \equiv \frac{\sqrt{(\hat{\omega}_*^T - \hat{\omega}_\alpha)^2 + \hat{\omega}_\psi^2 }}{\sqrt{\hat{\omega}_\alpha^2 + \hat{\omega}_\psi^2 } }.
\end{equation*}
Finally, one requires a boundary condition for the coordinate $\ell$ along the field line, which determines the behavior of trapped particles near the edges of the domain. We choose a periodic boundary condition for this longitudinal coordinate, so that any function $k(\ell)$ satisfies
\begin{equation*}
    k(\ell+L) = k(\ell).
\end{equation*}

\textit{Results.}
We have implemented a numerical scheme for calculating the AE and find that the calculations are very fast; in its current implementation the scheme requires only a few CPU minutes to obtain a sufficiently resolved result for a single configuration. We compare the AE of a tokamak and various stellarator devices with the saturated turbulent energy fluxes calculated by nonlinear flux-tube simulations using the \textsc{gene} code. \par  
We note that this code employs coordinates $(x, y)$ which are proportional to the gyroradius, namely
\[
    x = \frac{L_\text{ref}}{\rho_\text{ref}} \sqrt{\frac{\psi}{\psi_\text{tot}}}, \quad
        y =  \frac{L_\text{ref}}{\rho_\text{ref}} \frac{\alpha}{q_0} \sqrt{\frac{\psi_0}{\psi_\text{tot}}}, 
\]
where $q_0=\iota_0^{-1}$ denotes the inverse rotational transform at the center of the flux tube, $\rho_\text{ref}$ is some reference length scale of the order of the gyroradius, $L_\text{ref}$ is some global length scale which is typically taken to be the minor radius, and $\psi_\text{tot}$ is the total toroidal magnetic flux passing through the last closed flux surface. In terms of these coordinates, we take the characteristic length scale over which energy is available to be $\Delta x  =\Delta y = q_0$.
We now compare the dimensionless AE, as defined in Eq. \eqref{eq:normalized-AE-final}, with the normalized saturated radial energy flux $\widehat{Q}_\text{sat}$ defined as
\begin{equation}
    \widehat{Q}_{\text{sat}} \equiv \int_{t_\text{sat}} Q_e(t) \frac{ dt }{t_\text{sat}} \Bigg/ \left( \left[\frac{\rho_\text{ref}}{L_\text{ref}} \right]^2 n_0 T_0^{3/2}\sqrt{\frac{1}{m}} \right).
\end{equation}
Here $t_\text{sat}$ denotes the time span in which the energy flux saturates. The instantaneous electron energy flux density in the radial direction $Q_e(t)$ is defined as
\begin{equation}
    \begin{aligned}
        {Q}_e &= \frac{1}{V} \int \epsilon (\delta f_e) (\mathbf{v}_E \cdot \hat{\nabla} x) \, d \mathbf{x},
        \label{eq:energy-flux-def}
    \end{aligned}
\end{equation}
where $V$ is the volume of the simulated domain, $\mathbf{v}_E$ is the $\mathbf{E}\times \mathbf{B}$ drift, $\hat{\nabla}$ is the normalized gradient operator $\hat{\nabla}=\rho_\text{ref}\nabla$, and $\delta f_e$ is the fluctuating part of the electron distribution function $f_e$, so that $f_e = f_{M} + \delta f_e$ \cite{Gorler2010MultiscaleMicroturbulence}. The simulation set encompasses computations of collisionless electrostatic turbulence. Most of these are TEM turbulence driven by a density gradient only, and constitute a subset of simulations recently described by Proll \textit{et al.} \cite{Proll2022TurbulenceGradient}. We have also included two simulations of TEM turbulence driven by an electron temperature gradient only \footnote{To prevent the electron temperature gradient mode from affecting the TEM turbulence in these simulations, the ratio of the electron over the ion temperature has been increased to $7$, and the smallest and largest resolved binormal wavenumber were chosen to be $0.05 \; k_y \rho_\text{ref}$ and $12.8 \; k_y \rho_\text{ref}$ (W7-X) / $2.4 \; \; k_y \rho_\text{ref}$ (DIII-D) respectively.}. In either case there are no ion-temperature-gradient-driven instabilities or turbulence, which would draw energy from ions rather than electrons and thus require a more complicated form of the AE. 
\par
\begin{figure}[!t]
    \centering
    \includegraphics{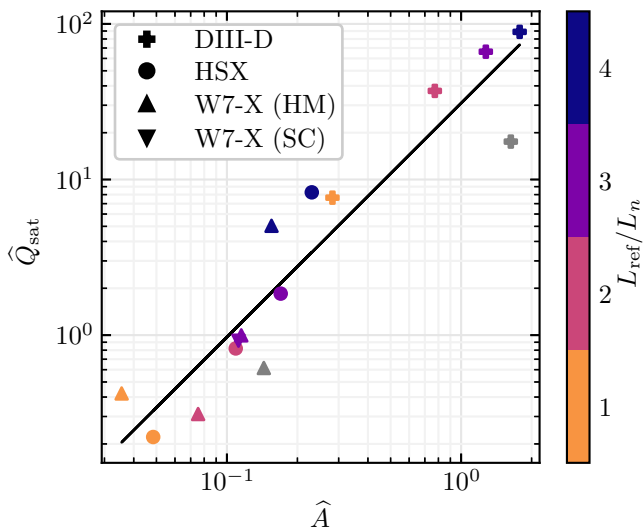}
    \caption{Correlation of the normalized AE and the nonlinear saturated radial energy flux. This is done with no electron temperature gradient and for different density gradients $L_\text{ref}/L_n$ indicated by the colorbar, and different devices. There are two gray points, which have no density gradient and a temperature gradient of $L_\text{ref}/L_T = 3$. The devices used in this analysis are the tokamak DIII-D, the Helically Symmetric eXperiment (HSX), and the Wendelstein 7-X (W7-X) stellarator in both high mirror (HM) and standard configuration (SC). The straight black line is the least-squares fit, which results in the power law $\ln Q_\text{sat} \propto (1.5 \pm 10\%) \ln A. $}
    \label{fig:scatter_plot_AE}
\end{figure}
The results of this comparison are plotted in Fig. \ref{fig:scatter_plot_AE}, which exhibits a strong correlation between the turbulent electron energy flux and AE over several orders of magnitude, in a tokamak and several stellarators, for various values of the density gradient, and for one value of the electron temperature gradient, $L_\text{ref}/L_T = 3$. Here $L_T= -T/(d T /d r)$ is the length scale of the electron temperature gradient, with $r$ being the minor radial coordinate. The density gradient has an analogous definition $L_\text{ref}/L_n$, with $L_n = - n/(dn/dr)$. We note that classical TEMs are thought to be largely absent in W7-X and ion-driven TEMs could instead be the dominant trapped-particle instability \cite{Proll2022TurbulenceGradient,Plunk2017CollisionlessMode}. These instabilities derive energy from the ions instead of electrons, but we nevertheless include these data points in Fig. \ref{fig:scatter_plot_AE}.
A power law is found by fitting a straight line to this doubly logarithmic plot, which gives
\begin{equation}
    Q_\text{sat} \propto A^{1.5 \pm 10\%}.
\end{equation}
This relation can be understood in the following manner. From the definition of the energy flux given in Eq. \eqref{eq:energy-flux-def}, we crudely estimate this flux as
\begin{equation*}
    Q_e \sim \sqrt{\langle \mathbf{v}_E^2 \rangle} \int \epsilon \delta f_e \, d \mathbf{x},
\end{equation*}
where the angular brackets denote a volume average.
The integral in this expression is bounded by the AE, and we thus set $\int \epsilon \delta f_e d \mathbf{x} \sim A$. The square of the drift velocity,
\begin{equation*}
    \langle \mathbf{v}_E^2 \rangle = \left\langle \left( \mathbf{E} \times \mathbf{B} / B^2 \right)^2 \right\rangle,
\end{equation*}
is proportional to the gyrokinetic energy of the electric field. Since the sum total of the thermal and this field energy is conserved \cite{Helander2017AvailablePlasmas}, the field energy is also bounded by the AE. Hence we estimate $\langle \mathbf{v}_E^2 \rangle \sim A$, which gives
\begin{equation}
    Q_e \propto A^{3/2},
\end{equation}
in agreement with the observed power law, within error bars.

\textit{Conclusions.}
As we have seen, it is possible to express the AE of trapped electrons in a slender flux tube of elliptical cross section in analytical form, which enables this quantity to be calculated efficiently. Since the AE represents a rigorous upper bound on the energy that can be converted into turbulent fluctuations, it is natural to compare it with the outcome of gyrokinetic simulations. The turbulent energy flux from the latter has thus been compared with the AE computed in flux tubes whose thickness perpendicular to the magnetic field is a fixed number of poloidal gyroradii. For TEM turbulence, one finds a strong correlation over several orders of magnitude and across a range of devices. The energy flux is found to be proportional to AE$^{3/2}$. 
\par 
These results are encouraging as they suggest a close connection between turbulent transport and AE, which could be used for quickly assessing confinement properties of  magnetic configurations without gyrokinetic simulations. The AE could serve as a proxy function in stellarator optimisation, both for new stellarator designs as well as existing devices, where one may adjust coil currents and plasma profiles to find AE-minimizing configurations. It is however not known whether the close correlation between AE and energy transport persists in plasmas with ion temperature-gradient-driven turbulence. Such turbulence draws energy from the ions, which are not constrained by the invariance of $\cal J$, a fact that needs to be accounted for in the calculation of the AE. \par
The authors are grateful for the valuable discussions with J. Ball, T. G\"orler, M.J. Pueschel, and M.J. Gerard. This work was partly supported by a grant from the Simons Foundation (560651, PH), and this publication is part of the project ``Shaping turbulence – building a framework for turbulence optimisation of fusion reactors'', with Project No. \texttt{OCENW.KLEIN.013} of the research program ``NWO Open Competition Domain Science'' which is financed by the Dutch Research Council (NWO). This work has been carried out within the framework of the EUROfusion Consortium, funded by the European Union via the Euratom Research and Training Program (Grant Agreement No. 101052200 — EUROfusion). Views and opinions expressed are however those of the author(s) only and do not necessarily reflect those of the European Union or the European Commission. Neither the European Union nor the European Commission can be held responsible for them.

\bibliography{references.bib}

\end{document}